\documentclass[12pt]{article}

\usepackage{geometry}
\usepackage{graphicx}
\usepackage{amsmath,amsfonts,amssymb,amscd,xspace}
\usepackage{longtable}
\usepackage{booktabs}
\usepackage[authoryear]{natbib}
\usepackage[english]{babel}
\usepackage{hyphenat}
\usepackage{xcolor}
\usepackage[utf8]{inputenc} 
\usepackage[toc,page]{appendix}
\usepackage{tikz}
\usepackage{lineno}
\usepackage{setspace}

\usepackage[pdfpagemode={UseOutlines},bookmarks=true,bookmarksopen=true,
bookmarksopenlevel=0,bookmarksnumbered=true,hypertexnames=false,
colorlinks,linkcolor={blue},citecolor={blue},urlcolor={red},
pdfstartview={FitV},unicode,breaklinks=true]{hyperref}

\newcommand{\blind}{0}

\newcommand{\E}{\ensuremath{\mathbb{E}}}

\newcommand{\exptiltedstand}{\overline{X}(\hat{\tau})}

\begin{document}

\def\spacingset#1{\renewcommand{\baselinestretch}%
{#1}\small\normalsize} \spacingset{1}


\if0\blind
{
  \title{\bf Saddlepoint-adjusted inversion of characteristic functions}
  \author{Berent \AA. S. Lunde\\
    Department of Mathematics and Physics, University of Stavanger\\
    and \\
    Tore S. Kleppe \\
    Department of Mathematics and Physics, University of Stavanger\\
	and \\
	Hans J. Skaug \\
	Department of Mathematics, University of Bergen}
  \maketitle
} \fi

\if1\blind
{
  \bigskip
  \bigskip
  \bigskip
  \begin{center}
    {\LARGE\bf Title}
\end{center}
  \medskip
} \fi

\bigskip
\begin{abstract}
For certain types of statistical models, the characteristic function (Fourier transform) is available in
closed form, whereas the probability density function has an intractable form, typically as an infinite sum of probability weighted densities.	Important such examples include solutions of stochastic differential equations with jumps, the Tweedie model, and Poisson mixture models. We propose a novel and general numerical method for retrieving the probability density function from the characteristic function, conditioned on the existence of the moment generating function.
Unlike methods based on direct application of quadrature to the inverse Fourier transform, the proposed method allows accurate evaluation of the log-probability density function arbitrarily far out in the tail. Moreover, unlike ordinary saddlepoint approximations, the proposed methodology is in principle exact modulus discretization and truncation error of quadrature applied to inversion in a high-density region. 
Owing to these properties, the proposed method is computationally stable and very accurate under log-likelihood optimisation. 
The method is illustrated for a normal variance-mean mixture, and in an application of maximum likelihood estimation to a jump diffusion model for financial data.
\end{abstract}

\noindent%
\vfill

\newpage
\spacingset{1.5} 
\section{Introduction}

This paper is motivated by the problem of making inference about parameters of statistical models, in which the probability density function is not readily available, but the characteristic function (CF) is.
Such models appear frequently when there are more than one source of randomness, which is usually the case in the real world;
\begin{itemize}
	\item[(1)] The compounded Poisson process and Tweedie model \citep{tweedie1984index, jorgensen1987exponential}, has a closed form characteristic function, which is usually applied when doing inference. 
	\item[(2)] The transition distributions of the class of affine jump diffusion models \citep{dps00} admit characteristic functions given in terms of the solution of certain ordinary differential equations, whereas the transition densities are typically not tractable.
	\item[(3)] The solution of non-linear stochastic differential equations (SDE) can be approximated well using Itô-Taylor expansions, for which the characteristic function can be derived \citep{preston2012approximation}.
	A further extension to the SDE is to add independent stochastic jumps that occur with an intensity at the given time. 
	The new sources of randomness, the counting process and the jump size, can then be easily added to the characteristic function of the approximate solution to the SDE \citep{zhang2016approximation}.
	\item[(4)] Other important examples include the non-central $\chi^2$, other Poisson mixtures, Normal mixtures and so on. 
\end{itemize}

Estimation methods based on the CF (or more generally, when the density/mass function is unavailable) can roughly be classified in two groups, based on whether or not the inverse Fourier transform (IFT) is used to compute likelihoods.
Popular methods in the latter group include (Generalized) Method of Moments (MM), 
which seeks to match a finite number of empirical moments with those from the model \citep[see e.g.][and subsequent literature]{10.2307/1912775}. A somewhat similar approach is that of using the empirical characteristic function (ECF), where the idea is to minimize a measure of the difference between the model-implied and empirical characteristic functions \citep[see e.g.][and references therein]{yu2004empirical}.

The MM is computationally very efficient, but as only a finite number of moment conditions are considered, these methods may discard important information from the data.
Such information loss does not in theory occur for ECF, but in practice ECF estimators often turn out to have asymptotic inefficiencies (see e.g.~\citet[Section~4]{knight2002theory}).

\begin{figure}
	\centering
	\includegraphics[width=0.45\textwidth,height=5cm]{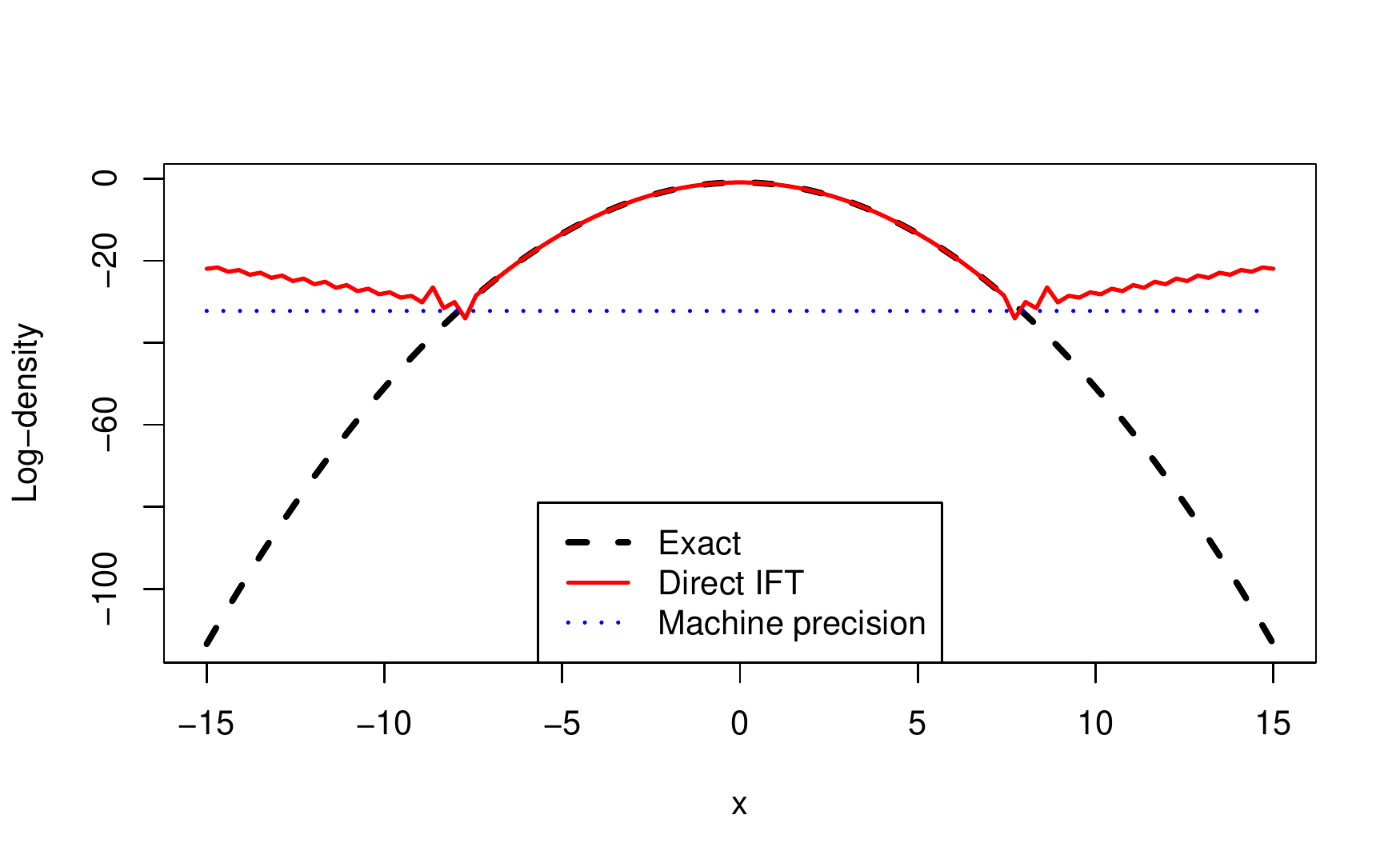}
	\caption{Logarithm of the $N(0,1)$ density calculated using quadrature-based direct inversion (Direct IFT), along with the exact log-density. Also indicated is $\log(1.0\times 10^{-14})$.  }
	\label{fig:tail_exact_spa}
\end{figure}

The former group of methods, e.g.~fully likelihood-based methods such as maximum 
likelihood estimation and Bayesian methods, approximates densities by evaluating IFTs 
numerically. 
Direct integration techniques, such as quadrature, will, as illustrated in 
Figure~\ref{fig:tail_exact_spa}, suffer from numerical inaccuracies 
that are non-trivial on log-scale in low-density regions. Such problems typically occur when the value of the true density 
is smaller than around $1.0\times 10^{-14}$, when using double numerics. This 
is related to the fact that the integrand of the IFT integral takes values $O(1)$. Consequently the
weighted sum of integrand evaluations constituting the quadrature approximation will be 
represented by a floating point number with integer exponent close to 0 \citep[see 
e.g.][Section 1.3, for a discussion of how floating point numbers are represented]{pressnumerical}. The spacing between representable numbers in such a floating 
point representation (typically on the order $1.0 \times 10^{-16}$ for exponent close to 
0) produces a theoretical lower bound on the magnitude of density values that can be 
accurately represented. However, from experience, accurate calculation of log-densities 
via quadrature approximations to the IFT typically fails when the density takes values
a few orders of magnitude higher than this bound. 

For likelihood-based estimation purposes, it is typically very important that log-densities can be evaluated in a stable manner, even far out in the tails. 
Therefore, density retrieval for estimation purposes will often involve rescaling properties of the density, such 
that the IFT is done in a high-density region and then scaled to the point of evaluation.
For example, for the mentioned Tweedie model, \citet{dunn2008evaluation} elaborate on density retrieval by the IFT using numerical integration. To bypass the problems in the tail, an analytically tractable rescaling property is stated and applied. Moreover, it is commented that satisfactory accuracy is only attained by utilizing this property. 
Unfortunately, such rescaling properties are not always evident for a general model.

An alternative to direct inversion is the saddlepoint approximation (SPA) 
\citep{daniels1954saddlepoint, barndorff1979edgeworth, butler2007saddlepoint}.
The SPA often admits fast computation, is highly automatic and allows for well-scaled
numerics when evaluating log-density approximations. 
In addition, the SPA has been shown to be accurate in the tails of the distribution 
\citep[see][]{butler2007saddlepoint}.
However, the SPA suffers some drawbacks in the context of likelihood estimation.
Most notably is that it in general does not integrate to 1 for non-Gaussian models, with values often $<1$ and thus, as a given parameter space may hold points that give Gaussian models where the SPA \textit{is} exact, may bias SPA-based estimation towards regions of the parameter space where the integral over the SPA is highest. 
Further, the SPA is often unimodal, and may thus produce an inaccurate approximation of e.g.~mixtures, 
which is a typical case in which the characteristic function is readily available 
but the density is not.

To bypass some of these problems, the SPA can be improved upon by employing a non-Gaussian 
base \citep{wood1993saddlepoint}.
\citet{ai2006saddlepoint} considers this approach for Markov processes, including jump 
diffusions. 
However, non-Gaussian based SPA requires the user to select a base-distribution with 
a density that in some sense is similar to the target density, 
and therefore may require bespoke implementations in each instance. 
\citet{kleppe2008building} introduce a method for choosing this base-distribution
automatically for the purpose of high-dimensional inversion under a
latent-variable framework. 

In this paper, we propose to precondition the integrand in univariate inversion problems
so that quadrature-based inversion is as numerically stable as possible.  This 
is done by inverting a standardised and exponentially tilted 
\citep{barndorff1979edgeworth} version of the target density, so that the inversion is
done in a high-density region. The method is in large parts automatic, and in particular does 
not rely on known rescaling properties of the target density or the elicitation of a
non-Gaussian base distribution.
Tilting is equivalent to deforming the contour of the inversion integral, and is a known technique in the applied mathematics literature for this type of problem \citep{cohen2007numerical}, so is rescaling.
Our approach is similar to that taken of \citet{strawderman2004computing}, which develops an automatic algorithm for inversion to the cumulative density function, albeit the saddlepoint is different than the saddlepoint employed as the tilt in this paper. \citet{strawderman2004computing} gives much attention to the numerical evaluation of the resulting inversion integral, and these considerations apply equally to the method we propose.

The rest of the paper is laid out as follows: 
the proposed method and its implementation are outlined in section \ref{sec: saddlepoint adjusted inversion}. In section \ref{sec: illustrations}, the proposed methodology is 
illustrated and contrasted to alternatives using the normal inverse Gaussian distribution
and the Merton jump diffusion models as example models. 
Conclusion and comments follow in section \ref{sec: discussion}.

\section{Methodology}
\label{sec: saddlepoint adjusted inversion}
This section provides some background and subsequently derives the saddlepoint adjusted 
IFT method proposed here.

\subsection{Background}
\label{subsec: direct ift}
Suppose that the strictly continuous random variable $X\in \mathbb{R}$ has probability density 
function $p_X(x)$ and CF $\varphi_X(s)=E_X(\exp(i s X))$, where $i=\sqrt{-1}$. 
As the CF may be considered as a
Fourier transform of $p_X(x)$, the density can be recovered from $\varphi_X(s)$ via an 
IFT: 
\begin{align}\label{eq: inverse fourier}
p_X(x) = \frac{1}{2\pi} \int_{-\infty}^\infty
\varphi_X(s)e^{-isx} ds.
\end{align}
For numerical evaluation, (\ref{eq: inverse fourier}) can be simplified by utilizing the Hermitian property of $\varphi_X(s)$:
\begin{equation}
p_X(x) 
= \frac{1}{\pi} \int_{0}^\infty \text{Re}\left[\varphi_X(s)e^{-isx} \right] ds.\label{eq:inv_real}
\end{equation}
As explained in the introduction and illustrated in Figure \ref{fig:tail_exact_spa}, 
computing the log-density $\log p_X(x)$ by taking the logarithm of the output of a 
quadrature method applied to either (\ref{eq: inverse fourier}) or (\ref{eq:inv_real}) is 
problematic when the value of $p_X(x)$ is small. Indeed, from
Figure \ref{fig:tail_exact_spa} it is seen that evaluation of the $N(0,1)$ log-density 
based on direct inversion fails around $p_X=\log(1.0\times 10^{-14})$ (indicated by the 
horizontal line).

\subsection{Saddlepoint adjusted IFT}
In order to avoid such numerical problems when evaluating log-densities, this section 
introduces a general method of preconditioning the integration problem. The overarching 
idea is to ensure that every numerical IFT approximates the density of a unit variance 
random variable at its mean. Modulus strongly pathological cases, this
approach should ensure that the value of the numerical Fourier transform is $O(1)$.

Again, suppose $X$ is the random variable of interest. Provided it exits, the cumulant
generating function (CGF) $K_X(t)$ is defined as $K_X(t)=\log\{E(\exp(tX))\}$, for values of $t\in \Omega$ where $\Omega = \{t \;:\; E(\exp(tX))<\infty\}$. In particular, the CGF may be recovered from the CF 
via
\[
K_X(t)=\log \{\varphi_X(-i t) \}.
\]
Suppose one wishes to evaluate $p_X$ at some point, say $x_0$.
A general and analytically tractable
rescaling of the original density $p_X$ is obtained by first introducing an exponentially 
tilted \citep[see e.g.][Section 2.4.5]{butler2007saddlepoint} version of $X$, say 
$X(\tau)$, where $\;\tau \in \Omega$ is the tilt parameter. The tilted random variable $X(\tau)$ has density
\begin{equation}
p_{X(\tau)}(x) = \exp(\tau x - K_X(\tau))p_X(x).\label{eq:tilted_pdf}
\end{equation}
Notice in particular that the original random variable $X$ is obtained for $\tau=0$. 
Straightforward calculations yield that the CGF associated with $X(\tau)$ is given as
\begin{align}
K_{X(\tau)} (t) 
= K_X (t+\tau) - K_X(\tau).\label{eq:tilted_CGF}
\end{align}
Now, in order for the evaluation of $p_{X(\tau)}$ to happen in a high-density region, 
the tilt parameter $\tau=\hat \tau$ is chosen so that $E(X(\hat \tau)) = x_0$. 
As $E(X(\tau)) = K^\prime_{X}(\tau)$, one obtains the saddlepoint equation  
\begin{equation}
K^\prime_{X}(\hat \tau) = x_0,\label{eq:saddlepoint_eq}
\end{equation} 
for $\hat \tau$, where $K^\prime_{X}$ denotes the derivative of $K_{X}$.

A further standardization is then introduced in order to ensure that the target for the 
numerical IFT also has unit variance (and thus under most circumstances has density values
$O(1)$ around the mean). This is achieved by introducing a standardized version of $X(\hat \tau)$,
\begin{equation}
\bar X(\hat \tau) = \frac{X(\hat \tau)-x_0}{\sqrt{Var(X(\hat \tau))}},\label{eq:stand_tilted}
\end{equation}
where $Var(X(\hat \tau)) = K^{\prime \prime}_{X}(\hat \tau)$, so that
\begin{equation}
p_{X(\hat \tau)}(x_0)=\frac{p_{\bar X(\hat \tau)}(0)}{\sqrt{K^{\prime \prime}_{X}(\hat \tau)}}. \label{eq:stand_tilted_pdf}
\end{equation}
Combining (\ref{eq:tilted_pdf}) and (\ref{eq:stand_tilted_pdf}) results in the 
preconditioning formula used throughout this text, namely
\begin{equation}
p_X(x_0) = \exp( K_X(\hat \tau) - \hat \tau x_0 ) \frac{p_{\bar X(\hat \tau)}(0)}{\sqrt{K^{\prime \prime}_{X}(\hat \tau)}}.\label{eq:spa_adjust_density}
\end{equation}
Since (\ref{eq:stand_tilted}) represents an affine transformation of $X(\hat \tau)$, the
CF and CGF of $\bar X(\hat \tau)$ are also easily found to be
\begin{eqnarray*}
	\varphi_{\exptiltedstand}(s) &=& \exp(K_{\exptiltedstand}(i s)),\\
	K_{\exptiltedstand}(t) &=& -K_X(\hat{\tau}) - \frac{t x_0}{\sqrt{K_X''(\hat{\tau})}}\\
	& & + K_X\left(
	\frac{t}{\sqrt{K_X''(\hat{\tau})}}+\hat{\tau}
	\right),
\end{eqnarray*}
and thus density of $\bar X(\hat \tau)$ evaluated at zero can be calculated as
\begin{equation}
p_{\bar X(\hat \tau)}(0) 
= \frac{1}{\pi}\int_0^\infty \text{Re}[\varphi_{\exptiltedstand}(s)]ds.\label{eq:p_bar_zero}
\end{equation}
\begin{figure*}
	\centering
	\includegraphics[scale=0.5]{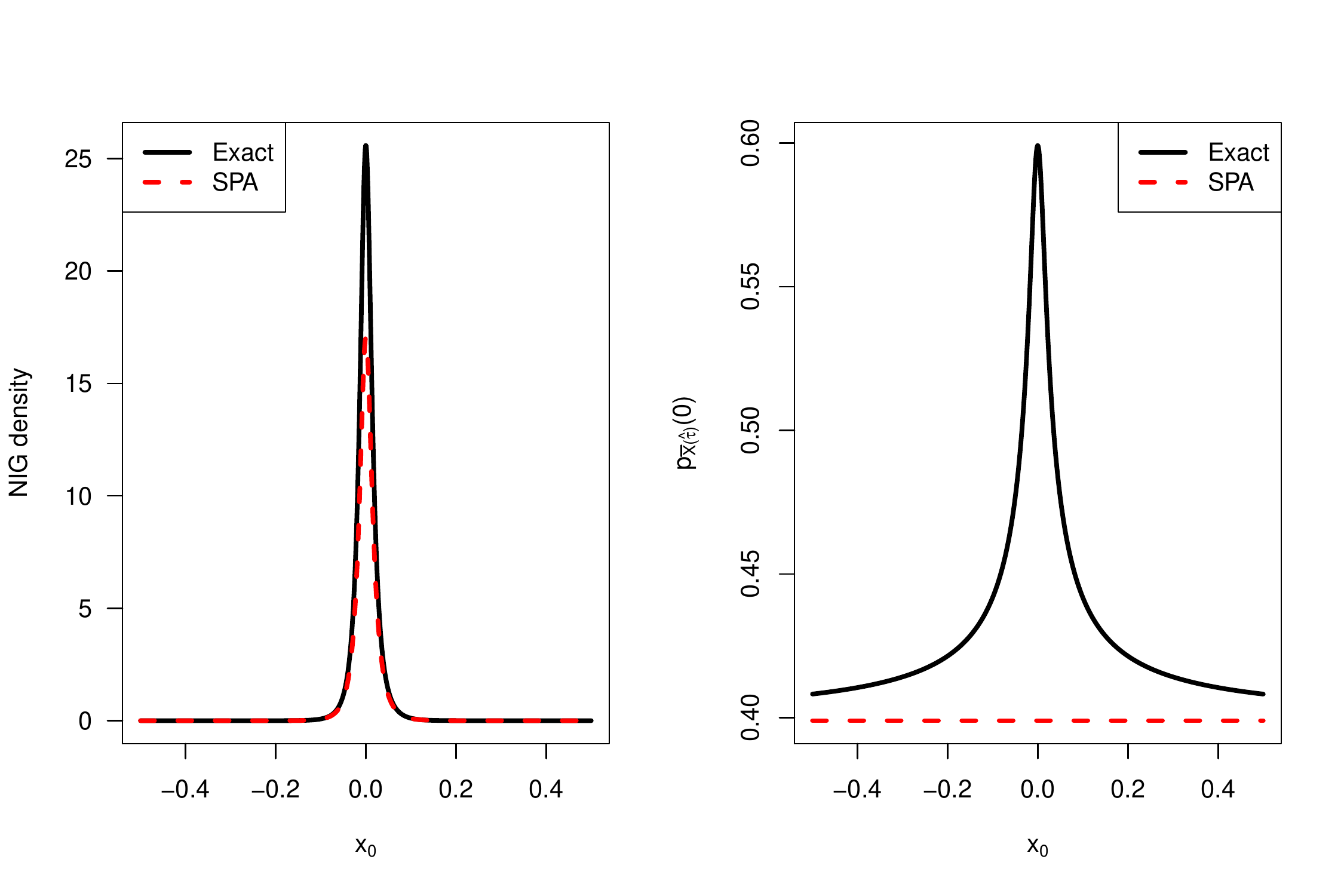}
	\caption{\label{fig:nig_exact_vs_spa}
		The normal inverse Gaussian (NIG) density 
		along with its SPA (left panel). The right panel shows $p_{\bar X(\hat \tau)}(0)$ (Exact) used in the proposed 
		methodology, together with the asymptote $(2\pi)^{-1/2}$ (SPA). The parameter values $\chi=0.0003$, $\psi=1000$, $\mu=-0.0003$, and $\gamma=2$ 
		in~(\protect\ref{NIG_10}) were applied in both plots. 
	}
\end{figure*}
Before proceeding, notice that the conventional SPA is obtained
by substituting $p_{\bar X(\hat \tau)}(0)$ in~(\ref{eq:spa_adjust_density}) with 
the $N(0,1)$ density evaluated at $0$, namely $(2\pi)^{-1/2}$. 
Thus, it follows from results on tail-exactness of the SPA for densities that are log-concave~ 
\citep{barndorff1999tail} that
also $p_{\bar X(\hat \tau)}(0)$ must converge to $(2\pi)^{-1/2}$ in the 
tails of $p_X$. In the high-density regions of $p_X$ on the other hand, $p_{\bar X(\hat \tau)}(0)$ typically takes values somewhat higher than $(2\pi)^{-1/2}$.
These properties
are illustrated for a normal inverse Gaussian distribution (to be discussed in more detail shortly) in Figure~\ref{fig:nig_exact_vs_spa}.
Note that $p_{\bar X(\hat \tau)}(0)$ approaches $(2\pi)^{-1/2}$ in the tails of the
distribution, suggesting that the SPA has the tail exactness property for the normal inverse Gaussian distribution.
Moreover, notice that $p_{\bar X(\hat \tau)}(0)$ remains well scaled $O(1)$ across the support of the density, and is
therefore easy to approximate using quadrature.

\subsection{Implementation}
\label{sec: implementation}
\begin{figure}
	\centering
	\includegraphics[scale=0.45]{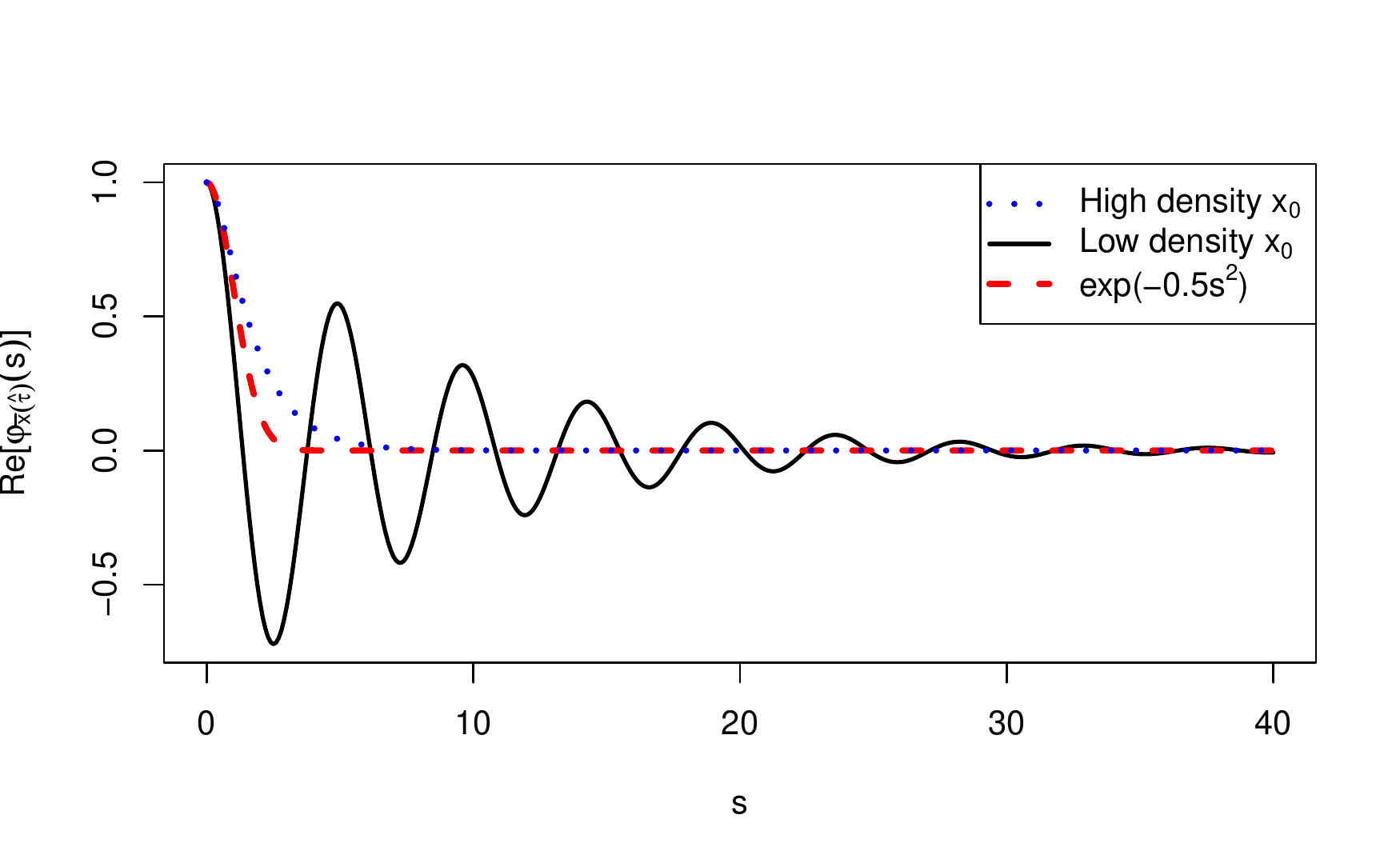}
	\caption{\label{fig:charfun_integrands} Integrands of the latter representation of 
		(\ref{eq:p_bar_zero}) for the normal inverse Gaussian distribution.
		The parameters are the same as in Figure~\ref{fig:nig_exact_vs_spa}. 
		High density $x_0$ correspond to $x_0=E(X)$ and low density to $x_0=E(X)-8\sqrt{Var(X)}$. As a reference, also the the CF associated with the $\bar X(\hat \tau)$ obtained whenever $X$ is Gaussian, i.e.~$\exp(-s^2/2)$ is indicated. It is
		seen that the integrands typically take non-negligible values for larger values of $s$ for
		relative to the Gaussian case. 
	}
\end{figure}
Provided the CGF/CF of some distribution $X$, equations (\ref{eq:spa_adjust_density}) and 
(\ref{eq:p_bar_zero}) form the basis for implementing the proposed saddlepoint adjusted
IFT technique for evaluating (log-)densities at say $x_0$. 
Each evaluation involves the following steps:
\begin{enumerate}
	\item Obtain the saddlepoint $\hat \tau $ by solving (\ref{eq:saddlepoint_eq}). 
	Notice that $\hat \tau  = \arg \min_{t\in\Omega} K_X(t)-x_0 t$, where the objective 
	function is convex on $\Omega$. The convexity ensures both a unique such solution to
	(\ref{eq:saddlepoint_eq}), and also that Newton's method of optimization \citep[Chapter~9.4]{pressnumerical} may be used to obtain
	a rapidly- and stably converging solution.
	\item Approximate the latter integral of (\ref{eq:p_bar_zero}) by a quadrature approximation, say $\hat p_{\bar X(\hat \tau)}(0)$.
	\item Compute log-density approximation as
	\begin{align*}
	\log(p_X(x_0)) \approx &K_X(\hat \tau) - \hat \tau x_0 - \frac{1}{2}\log\left(K^{\prime \prime}_{X}(\hat \tau) \right)\\ &+ \log(\hat p_{\bar X(\hat \tau)}(0)).
	\end{align*}
\end{enumerate}

The saddlepoint adjusted IFT comes with some extra cost compared to the saddlepiont 
approximation and, depending on number of quadrature evaluation, potentially direct IFT. 
However, for the normal-inverse Gaussian model considered shortly, the location of a 
single saddlepoint $\hat \tau$ per evaluation is a minor part of the 
required CPU time.  This is typically the case even if the saddlepoint equation must be 
solved numerically, as is 
often the case for non-trivial models. 
Thus, of highest importance for good and 
robust performance is the selection of a quadrature rule for implementing point 2 above. 

Figure~\ref{fig:charfun_integrands} displays
integrands $\text{Re}[\varphi_{\exptiltedstand}(s)]$ for the normal inverse Gaussian 
distribution also considered in Figure~\ref{fig:nig_exact_vs_spa}. For values of $x_0$ in 
the high-density region of $X$, the integrand falls to zero rapidly, and the resulting 
integral may be accurately approximated using Gauss-Hermite quadrature 
\citep[p.~153]{pressnumerical}. On the other hand, is seen that the
integrand may take non-trivial values far from the origin when $x_0$ is in the tails
of $X$ (even if the resulting integral attains values close to $(2\pi)^{-1/2}$ \citep{barndorff1999tail}). 

In the present work, composite Simpson's quadrature with a fixed integration range
is used for the integration problem in point 2 above. This choice is made mainly for 
robustness and in order to obtain (log-)density approximations that are smooth functions
in the parameters.
This works fine for the selected illustrations, but more heavy-tailed distributions that have 
a slower convergence of $\lim_{s\to\infty}\text{Re}[\varphi_{\exptiltedstand}(s)]\to 0$, 
might severely suffer from truncation error.
\citet{abate1992fourier} argues for the even simpler Trapezoid method for inversion problems,
which' relation to the Poison summation formula may be used to control discretization error,
in combination with Euler-summation for convergence acceleration which targets truncation error. 
\citet{strawderman2004computing} extensively discuss the issues of discretization and truncation 
error, and suggests Wynn's epsilon-method for convergence acceleration over Euler-summation.
A more adaptive selection of integration range that is also smooth
in parameters, type of convergence acceleration, and also choosing integration rules that account for the potentially
oscillating nature of the integrand holds scope for future research. Notice, however, that
such adaptive integration would not alone solve the problems with log-density evaluation
for direct IFT as illustrated in Figure~\ref{fig:tail_exact_spa}.

Notice, for comparison, that renormalised (via numerical integration) 
SPAs require the solution of many saddlepoint
equations, while at the same time, introduce non-vanishing approximation errors and loses tail 
exactness. Thus, due both to a higher computational cost and non-vanishing errors,
renormalised SPAs are not considered further here.

Throughout this paper, all methods are implemented in C++ and run in R \citep{r2018project} using the RCPP
package \citep{eddelbuettel2011rcpp}. Exact gradients of log-likelihood functions were
obtained using the automatic differentiation (AD) library Adept \citep{hogan2014fast}. 
All derivatives of the CGF are hand-coded in the present work, but this process may also 
be automated using a tool that allows for nested computation of derivatives such as 
TMB \citep{kristensen2016tmb}.

\section{Illustrations}
\label{sec: illustrations}

The focus of this section is to highlight several properties of the saddlepoint adjusted IFT method, and to contrast these to the SPA and direct IFT.
The included target distributions hold Gaussian solutions as special cases in their respective parameter space, and thus in some part selected for their illustrative purposes with regards to the Gaussian bias in SPA based estimation.

\subsection{The Normal inverse Gaussian distribution}
Throughout this section, the normal-inverse Gaussian (NIG) \citep[see e.g.][]{mcneilquantitative} is used as a
test case since this distribution admit exact evaluation of the density.
The NIG distribution is defined as a normal-variance mixture,
\begin{align}
X = \mu+\gamma W + \sqrt{W}Z, \; Z\sim N(0,1),
\label{NIG_10}
\end{align}
where $W$ has an inverse Gaussian distribution with a parametrization corresponding to 
\begin{align*}
p_W(w)\propto w^{-3/2}\exp(-(\chi/w+\psi w)/2),\\
E(W)=\sqrt{\chi/\psi}
\text{ and }
Var(W)=(\sqrt{\chi/\psi})/\psi.
\end{align*}
The marginal density of $X$ is expressible only in terms of the modified Bessel function of the third kind $K_\lambda$:
\begin{align*}
p_X(x)=&\frac{\sqrt{\chi (\psi+\gamma^2)} K_{-1}\left( \sqrt{(\chi + (x-\mu)^2)(\psi+\gamma^2)}  \right)  }
{\pi \sqrt{\chi+(x-\mu)^2}}\\ 
&\times e^{\sqrt{\chi\psi}+ (x-\mu)\gamma}.
\end{align*}
On the other hand, the CGF is highly tractable:
\begin{align*}
K_X(t) = 	s\mu + \sqrt{\chi}\left(\sqrt{\psi}-\sqrt{\psi-s^2 - 2s\gamma}\right).
\end{align*}
In particular, based on the CGF, one obtains $E(X)=\mu + \gamma \sqrt{\chi/\psi}$ and
$Var(X)= \sqrt{\chi}\psi^{-3/2}(\gamma^2-\psi)$.

In order to implement saddlepoint adjusted IFT, a quadrature scheme must be chosen to approximate $p_{\bar X(\hat \tau)}(0) $ as given in the latter representation of (\ref{eq:p_bar_zero}). Here, Simpson's quadrature based on 512 equidistant evaluations in the interval $s\in [0,100]$ were used throughout.
The interval and number of evaluations were chosen to be highly robust for a wide range of
parameters and evaluation points $x_0$. Notice, that direct IFT on the other hand requires
more manual tuning depending on parametrisation.

\begin{figure*}
	\centering
	\includegraphics[scale=0.45]{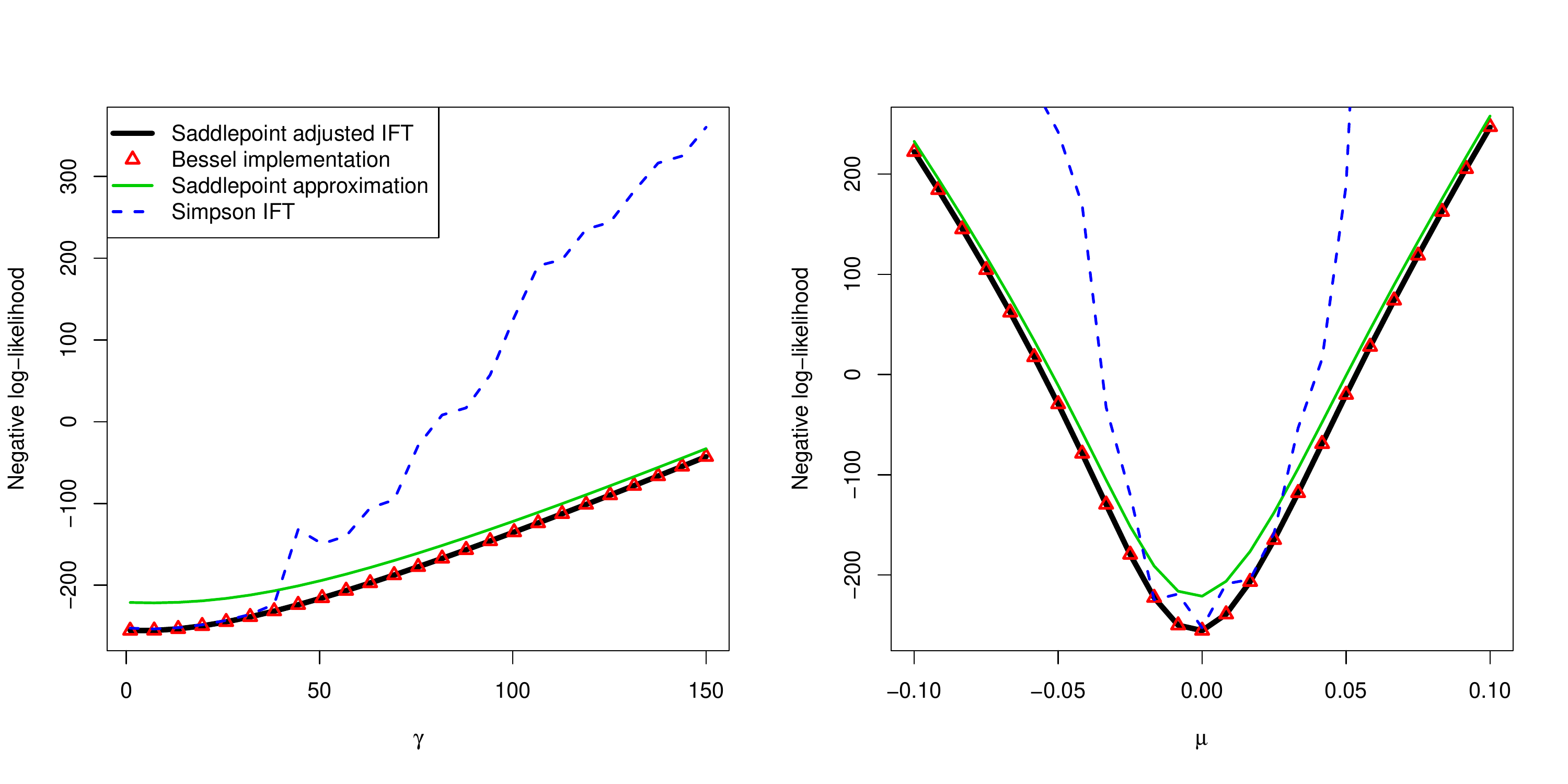} 
	\caption{ \label{fig:nll_nig_inversions }
		Approximate (negative) log-likelihood profiles based on 100 observations simulated from a NIG distribution with parameter $\chi=0.0003$, $\psi=1000$, $\mu=-0.0003$, 
		and $\gamma=2$. The left panel shows log-likelihoods plotted as a function of $\gamma$ while keeping the remaining parameters fixed at their ``true" values. 
		The right panel shows a similar plot with varying $\mu$. 
	}
\end{figure*}

To illustrate how the saddlepoint adjusted IFT performs
under log-likelihood-based estimation, 100 observations from a NIG distribution with 
parameters $\chi=0.0003$, $\psi=1000$, $\mu=-0.0003$, and $\gamma=2$ were simulated. These parameter values correspond to $E(X)\approx 0.0008$ and $Var(X)\approx 0.023^2$, 
which are typical of financial returns.
The log-likelihood, approximated by saddlepoint adjusted IFT, the SPA, the direct IFT, and 
using the native R Bessel implementation (regarded as being exact), are plotted in Figure~\ref{fig:nll_nig_inversions } as functions of $\gamma$ and $\mu$, respectively, while keeping the remaining parameters fixed at the values used for simulation. 
In both cases it is seen that the direct IFT based on Simpson's method produces unreliable results for parameter settings far from the ``true" parameters. 
Moreover, the SPA produces a log-likelihood approximation which deviates from the true log-likelihood (Bessel implementation) with a parameter-dependent amount. 
Hence, the SPA based log-likelihood approximation is likely to result in biases relative to the exact maximum likelihood estimator. 
Finally, it is seen that the saddlepoint adjusted IFT based approximation is indistinguishable from the true log-likelihood. 

To avoid numerical problems resulting from non-positive density approximations, the direct IFT 
log-likelihood was 
implemented as $\log(\max(1.0e-14,\hat I))$ where $\hat I$ is the quadrature-based 
approximation of (\ref{eq:inv_real}). Such non-positive density approximations are obtained as
a consequence of a highly oscillating integrand that takes values far larger than the value
of the integral itself. Simpson's quadrature, with 512 function evaluations over the integration range  $s\in [0,150]$,
was used. Nevertheless, Figure~\ref{fig:nll_nig_inversions } shows that direct IFT 
produces erratic behaviour for parameters far from those used in the
simulation. When increasing $\gamma$ (left panel) the distribution of $X$
changes from being rather symmetric around $\gamma=2$ to being highly skewed with a heavy 
right tail and a very thin left tail around $\gamma=150$.
Thus, in this latter case, some of the simulated observations will have  
very small density values, which is problematic for direct IFT, as demonstrated in Figure~\ref{fig:tail_exact_spa}.
Similar reasoning holds for the right panel, where the location parameter $\mu$ is is varied by around 4 standard deviations in either direction 
from the ``true" $\mu$, and in this case the majority of the simulated data are in the far 
tails at either extreme of the plot.

As is also seen in Figure~\ref{fig:nll_nig_inversions }, the SPA produces approximate
log-likelihoods that deviate by a parameter-dependent offset from the exact 
log-likelihood. This behaviour is related to the fact that the SPA typically does not 
integrate to 1 (typically $<1$), and that the appropriate normalisation factor of the SPA 
is parameter dependent. However, the SPA is exact in the Gaussian case, and it is therefore
often seen \citep[see e.g.][]{kleppe2008building} that SPA-based parameter estimates are
biased towards the ``Gaussian part" of the parameter space in models that nest Gaussian distributions. For the NIG
distribution, a $N(\mu+\gamma\sigma^2,\sigma^2)$ distribution obtains when 
$E(W)\rightarrow \sigma^2$ and $Var(W)\rightarrow 0$ (e.g.~when $\chi=\sigma^4 \psi$ and $\psi \rightarrow \infty$). 

To illustrate this effect, we fixed parameters $\mu=\gamma=0$ and chose $\chi=\psi$ so that $\E[W]=1$. The remaining free parameter $\theta=1/\psi=Var(W)$ controls the variance of $W$, and thus the
deviation from normality, with $X\rightarrow N(0,1)$ as $\theta \rightarrow 0$.
We then computed estimators of $\theta$ by numerically minimizing the relative Kullback-Leibler divergence to obtain asymptotic maximum likelihood estimators,
\begin{align}
\hat{\theta}(\theta_0) = \arg\min_{\theta} 
\left\{
- \int \log \{\tilde{p}_X(x;\theta)\} p_X(x;\theta_0) dx
\right\},
\label{eq:asymp_MLE}
\end{align}
for different settings of the inverse Gaussian variance $\theta_0$. Here, 
$\tilde{p}$
is either the SPA or the saddlepoint adjusted IFT.
The integral in the objective function of
(\ref{eq:asymp_MLE}) was resolved using quadrature with 200 evaluations on
an interval centered at the mean and spanning 12 standard deviations.
The results showed that for the SPA we have $\hat{\theta}(\theta_0)=0$, i.e.~a Gaussian distribution, for all values of $\theta_0$. For the saddlepoint adjusted 
IFT, on the other hand,  $\hat{\theta}(\theta_0)$ is 
indistinguishable from $\theta_0$, i.e.~the estimator is asymptotically unbiased.

\subsection{Application to real data: Merton jump diffusion}
\label{sec: applications to real data}

\begin{figure*}
	\centering
	\includegraphics[scale=0.5]{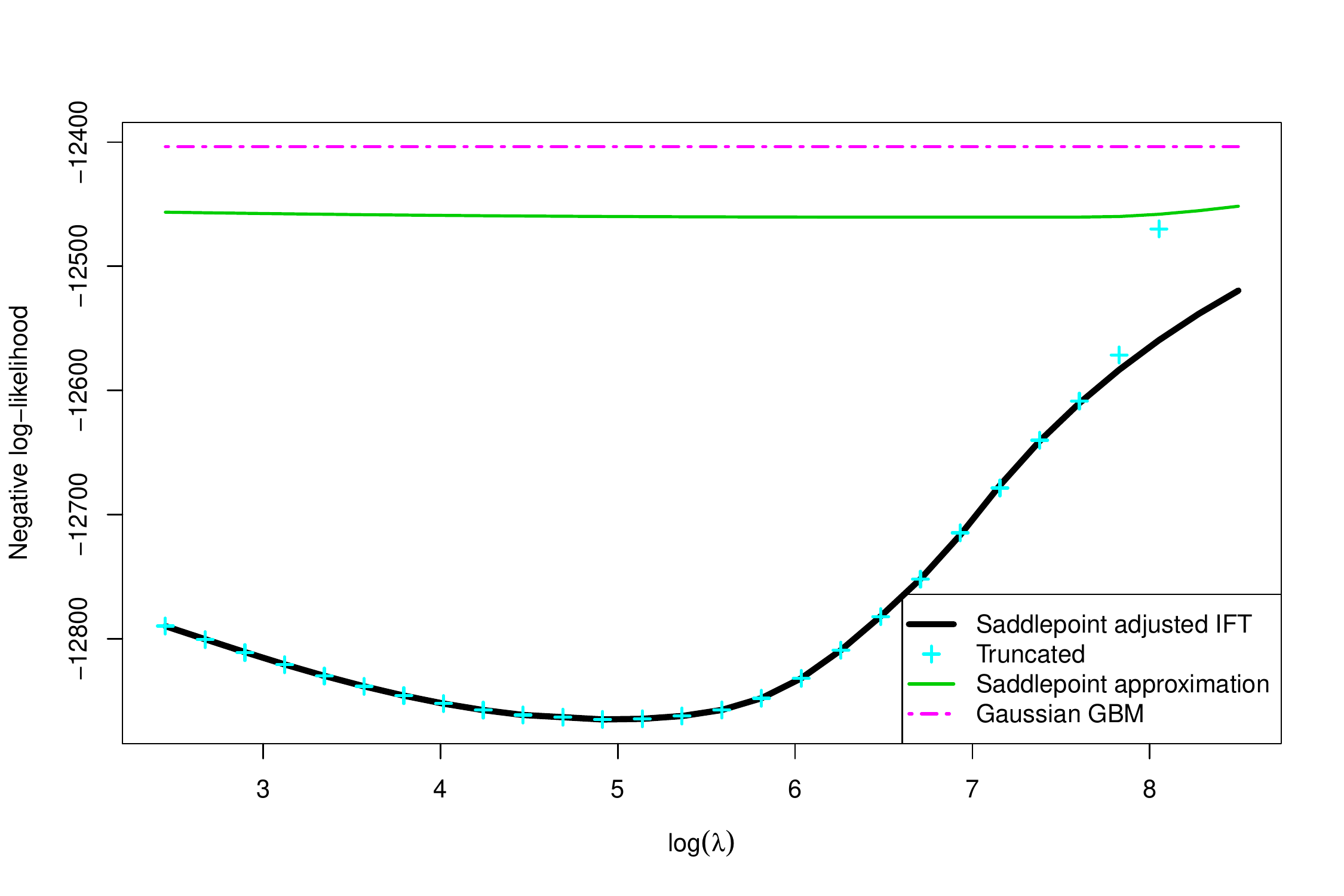} 
	\caption{
		Profile negative log-likelihoods of the Merton Jump Diffusion model versus fixed logarithmic jump-intensities, $\lambda$, fitted to DJIA daily stock-market data from 01.01.2000 until 01.01.2018.
		The evaluations referred to as ``Truncated" are obtained by summing conditional probabilities in \eqref{eg:Merton_truncated}, until either the jump probability is less than $1.0\times 10^{-14}$ or reaching 20 jumps.
		Saddlepoint adjusted IFT uses Simpson's quadrature with 128 equidistant steps over $s\in \left[0, 16\right]$.
		Direct inversion was deemed infeasible, as it severely suffers from the numerical problems discussed in this article. 
	}
	\label{fig:profile_ll_mjd_djia}
\end{figure*}


\begin{table*}
	\centering
	\begin{tabular}{@{}ccccc@{}}
		\toprule
		& \multicolumn{4}{c}{Methods}\\
		\cmidrule(r){2-5}
		& SPI & Truncated & SPA & GBM \\
		\cmidrule(r){2-2} \cmidrule(r){3-3} \cmidrule(r){4-4} \cmidrule{5-5}
		$r$ & 0.0445 (0.045) &  0.0445 (0.045) &  0.0432 (0.043) & 0.0461 (0.047)  \\
		$\log\sigma$ &   -2.41 (0.077) & -2.41 (0.076) &  -2.23 (1.11) & -1.67 (0.011)  \\
		$\log\lambda$ &  4.96 (0.18) & 4.96 (0.18) &   6.81 (2.36) & \\
		$\mu$ &  -0.00114 (0.00039) & -0.00114 (0.00039) &  -0.000703 (0.00065) & \\
		$\log\nu$ &   -4.32 (0.074) &  -4.32 (0.074) &  -5.4 (0.51) & \\
		\bottomrule
	\end{tabular}
	\caption{Likelihood estimates and standard deviations in parentheses for the MJD fitted to the DJIA daily stock-market data from 2000 until 2018, applying different methods for likelihood approximation. SPI refers to saddlepoint adjusted IFT.
		The settings for the methods are the same as those described in Figure~\protect\ref{fig:profile_ll_mjd_djia}.
		The inversion methods were implemented using first order AD data-types with Adept \citep{hogan2014fast}, and Hessian matrices could therefore be retrieved as finite-difference Jacobians to the exact gradients; these could in turn be used to calculate standard deviations (see for instance \citet{kristensen2016tmb} for details on such computations). The Hessian of the truncation method was retrieved via the \texttt{optim} function in R \citep{r2018project}.
	}
	\label{tab:mjd_par_est_real_data}
\end{table*}

This section considers the application of inversion techniques to likelihood optimisation 
based on real data.
Specifically, the Merton jump diffusion (MJD) \citep{merton1976option} for stock prices is 
considered. Under the MJD model, the dynamics of a stock price $S_t$ are described by the 
jump diffusion model
\begin{align}\label{sec:illustration:mjdsde}
\frac{dS_t}{S_{t^-}}=(r-\lambda k)dt + \sigma dW_t + (Y_t-1)dN_t,
\end{align}
where $r>0$ is the instantaneous expected return on the asset, $\sigma>0$ the 
instantaneous volatility (if a jump does not occur), and $k$ the expected relative jump 
size. $N_t$ denotes a Poisson process with intensity $\lambda > 0$, independent of the 
Brownian motion $W_t$.
$Y_t$ is the price jump size, meaning that, if a jump 
occurs at time $t$, the price jumps from $S_{t^-}$ to $Y_tS_{t^-}$. 
The jump sizes $Y_t$ are assumed to be log-normally distributed with $\log Y_t \sim 
N(\mu,\nu^2)$, and thus $k=e^{\mu+0.5\nu^2}-1$. 

The SDE \eqref{sec:illustration:mjdsde} can be solved to yield the following 
representation for logarithmic prices:
\begin{equation}
X_t=X_0+\left(r-\lambda k - \frac{\sigma^2}{2}\right)t + \sigma W_t 
+ \sum_{i=1}^{N_t}\log Y_i,\label{eq:X_rep_mixture}
\end{equation}
and thus the conditional CGF of $X_t|X_0$ is given as a sum of a normal CGF and a normal compounded 
Poisson CGF:
\begin{align}
K_{X_t|X_0}(s) =&
sX_0 + st\left(r-\lambda k - \frac{\sigma^2}{2}\right) +
\frac{s^2\sigma ^2 t}{2} \notag\\
&+
\lambda t \left(e^{s\mu+\frac{\nu^2s^2}{2}}-1\right). \label{eg:Merton_cgf}
\end{align}
The conditional CGF (\ref{eg:Merton_cgf}) does not appear to admit a closed form 
saddlepoint, $\hat{\tau}$, and thus numerical solution of (\ref{eq:saddlepoint_eq}) is 
required.

Notice that the Geometric Brownian Motion (GBM) processes for $S_t$ obtains as special
cases either for $\lambda=0$, or when $\mu=0$, $\lambda \rightarrow \infty$ and $\nu$ 
decays as $O(\lambda^{-1/2})$ or faster. Furthermore, $X_t|X_0$ is Gaussian
under the GBM model.

Note also that for the MJD, $X_t|(X_0,N_t)$ is Gaussian, and thus the exact transition
probability density is available as a Poisson mixture with Gaussian components: 
\begin{equation}
P(X_t|X_0) = \sum_{i=0}^{\infty}P(N_t = i)P(X_t|X_0,N_t=i). \label{eg:Merton_truncated}
\end{equation}
Specifically, $X_t|(X_0,N_t)$ is Gaussian with mean 
$X_0 + t\left(r-\lambda k - \frac{\sigma^2}{2}\right) + N_t\mu$
and variance
$\sigma^2 t + N_t\nu^2$.
As a 
reference for the saddlepoint adjusted IFT, we also consider an approximate log-likelihood function
(referred to as ``truncated"), based on truncating the Poisson mixture infinite sum 
representation either when the Poisson weights become smaller than 1.0e-14 or at 20 jumps
per transition. 

The MJD was applied to stock-market data from the Dow Jones Industrial Average index from 
01.01.2000 until 01.01.2018. A yearly time scale, and thus observations separated in time
with $t=1/252$, was used.
Maximum likelihood estimation based on saddlepoint adjusted IFT (using Simpson's 
quadrature with 128 equidistant evaluations over $s\in \left[0, 16\right]$), SPA and the 
truncated method were
considered. Direct IFT was deemed infeasible due to similar problems as discussed above.

Table \ref{tab:mjd_par_est_real_data} provides maximum likelihood estimates. First of all,
it is seen that the parameter estimates obtained using saddlepoint adjusted IFT and
the truncated method are close to indistinguishable. This observation suggests that the 
proposed methodology performs very well in also this situation, even when using a static 
integration rule. The parameter estimates suggest
rather frequent jumps at a rate of around 0.6 per day. The jumps have close to zero mean
and a standard deviation that is roughly double that of the diffusive part (i.e.~$\sigma 
\sqrt{t}$).

Also included in Table \ref{tab:mjd_par_est_real_data} are parameter estimates obtained
using a SPA-based log-likelihood approximation. This approximation favours a model with
smaller and more frequent jumps, which as discussed above, suggests a model with closer
to Gaussian transition distributions. Still, the model obtained using SPA-based
log-likelihood is well separated from the exactly Gaussian GBM.

Figure~\ref{fig:profile_ll_mjd_djia} presents negative profile log-likelihoods over 
$\log(\lambda)$ based on the different considered methods. It is seen that moderate
jump intensities, (say below $\log(\lambda)<7$, $P(N_t=20|\log(\lambda)=7)=3.1e-8$),
the truncated and saddlepoint adjusted IFT profile log-likelihoods coincide very
well, whereas for higher jump intensities, they diverge as the truncation of jump counts
starts taking effect. Though not particularly relevant for the data and model at hand 
here, these observations illustrate the utility of performing the mixing in 
transform-space, as the computational complexity remains the same for any jump intensity,
whereas summing many jumps in the Gaussian mixture representation may be become very
computationally expensive. 

Figure~\ref{fig:profile_ll_mjd_djia} also includes profile likelihoods for SPA-based 
approximate log-likelihood and the GBM model (invariant of $\lambda$). It is seen that
the approximate log-likelihood associated with the SPA is substantially closer to the
GBM. Notice also that for high jump intensities, the saddlepoint adjusted IFT and
SPA approach the GBM as close to Gaussian models are obtained for very high jump 
intensities are imposed. The truncated method, on the other hand, fails in representing
this effect for high jump intensities.

\section{Discussion}
\label{sec: discussion}

This paper proposes a new method for numerical inversion of characteristic functions, 
conditioned on the existence of the CGF.
The proposed method is very reliable for obtaining log-density values, even far out in the 
tails of the distribution. In particular, the method resolves numerical problems that may
occur using direct inverse Fourier transformation. Moreover, the method may be
seen as a way of substantially improving the accuracy of the classical saddlepoint 
approximation when applied in likelihood-based inference. 

Further work will involve more automatic rules for choosing integration 
ranges, under the constraint of producing smooth (in parameters) log-likelihood functions.
A further extension would be to consider the low- but multi-dimensional analogue of the 
proposed methodology. However, more work regarding how to implement the resulting multidimensional integral in point 
2 in section \ref{sec: implementation} must be carried out also in this case.

%
%
%
%
%
%

\bibliographystyle{Chicago}

\bibliography{Bibliography-SPI}

\begin{thebibliography}{}

\bibitem[\protect\citeauthoryear{Abate and Whitt}{Abate and
  Whitt}{1992}]{abate1992fourier}
Abate, J. and W.~Whitt (1992).
\newblock The fourier-series method for inverting transforms of probability
  distributions.
\newblock {\em Queueing systems\/}~{\em 10\/}(1-2), 5--87.

\bibitem[\protect\citeauthoryear{Ait-Sahalia and Yu}{Ait-Sahalia and
  Yu}{2006}]{ai2006saddlepoint}
Ait-Sahalia, Y. and J.~Yu (2006).
\newblock Saddlepoint approximations for continuous-time {Markov} processes.
\newblock {\em Journal of Econometrics\/}~{\em 134\/}(2), 507--551.

\bibitem[\protect\citeauthoryear{Barndorff-Nielsen and Cox}{Barndorff-Nielsen
  and Cox}{1979}]{barndorff1979edgeworth}
Barndorff-Nielsen, O. and D.~R. Cox (1979).
\newblock Edgeworth and saddle-point approximations with statistical
  applications.
\newblock {\em Journal of the Royal Statistical Society. Series B
  (Methodological)\/}, 279--312.

\bibitem[\protect\citeauthoryear{Barndorff{\hyp }Nielsen and
  Kluppelberg}{Barndorff{\hyp }Nielsen and
  Kluppelberg}{1999}]{barndorff1999tail}
Barndorff{\hyp }Nielsen, O.~E. and C.~Kluppelberg (1999).
\newblock Tail exactness of multivariate saddlepoint approximations.
\newblock {\em Scandinavian journal of statistics\/}~{\em 26\/}(2), 253--264.

\bibitem[\protect\citeauthoryear{Butler}{Butler}{2007}]{butler2007saddlepoint}
Butler, R.~W. (2007).
\newblock {\em Saddlepoint approximations with applications}.
\newblock Cambridge University Press.

\bibitem[\protect\citeauthoryear{Cohen}{Cohen}{2007}]{cohen2007numerical}
Cohen, A.~M. (2007).
\newblock {\em Numerical methods for Laplace transform inversion}, Volume~5.
\newblock Springer Science \& Business Media.

\bibitem[\protect\citeauthoryear{Daniels}{Daniels}{1954}]{daniels1954saddlepoint}
Daniels, H.~E. (1954).
\newblock Saddlepoint approximations in statistics.
\newblock {\em The Annals of Mathematical Statistics\/}, 631--650.

\bibitem[\protect\citeauthoryear{Duffie, Pan, and Singleton}{Duffie
  et~al.}{2000}]{dps00}
Duffie, D., J.~Pan, and K.~Singleton (2000).
\newblock Transform analysis and asset pricing for affine jump-diffusions.
\newblock {\em Econometrica\/}~{\em 68\/}(6), 1343--1376.

\bibitem[\protect\citeauthoryear{Dunn and Smyth}{Dunn and
  Smyth}{2008}]{dunn2008evaluation}
Dunn, P.~K. and G.~K. Smyth (2008).
\newblock Evaluation of {T}weedie exponential dispersion model densities by
  {F}ourier inversion.
\newblock {\em Statistics and Computing\/}~{\em 18\/}(1), 73--86.

\bibitem[\protect\citeauthoryear{Eddelbuettel and Fran\c{c}ois}{Eddelbuettel
  and Fran\c{c}ois}{2011}]{eddelbuettel2011rcpp}
Eddelbuettel, D. and R.~Fran\c{c}ois (2011).
\newblock {Rcpp}: Seamless {R} and {C++} integration.
\newblock {\em Journal of Statistical Software\/}~{\em 40\/}(8), 1--18.

\bibitem[\protect\citeauthoryear{Hansen}{Hansen}{1982}]{10.2307/1912775}
Hansen, L.~P. (1982).
\newblock Large sample properties of generalized method of moments estimators.
\newblock {\em Econometrica\/}~{\em 50\/}(4), 1029--1054.

\bibitem[\protect\citeauthoryear{Hogan}{Hogan}{2014}]{hogan2014fast}
Hogan, R.~J. (2014).
\newblock Fast reverse-mode automatic differentiation using expression
  templates in {C}++.
\newblock {\em ACM Transactions on Mathematical Software (TOMS)\/}~{\em
  40\/}(4), 26.

\bibitem[\protect\citeauthoryear{Jorgensen}{Jorgensen}{1987}]{jorgensen1987exponential}
Jorgensen, B. (1987).
\newblock Exponential dispersion models.
\newblock {\em Journal of the Royal Statistical Society. Series B
  (Methodological)\/}, 127--162.

\bibitem[\protect\citeauthoryear{Kleppe and Skaug}{Kleppe and
  Skaug}{2008}]{kleppe2008building}
Kleppe, T.~S. and H.~J. Skaug (2008).
\newblock Building and fitting non-{Gaussian} latent variable models via the
  moment-generating function.
\newblock {\em Scandinavian Journal of Statistics\/}~{\em 35\/}(4), 664--676.

\bibitem[\protect\citeauthoryear{{K}night, {S}atchell, and {Y}u}{{K}night
  et~al.}{2002}]{knight2002theory}
{K}night, J.~L., S.~E. {S}atchell, and J.~{Y}u (2002).
\newblock Theory \& methods: Estimation of the stochastic volatility model by
  the empirical characteristic function method.
\newblock {\em Australian \& New Zealand Journal of Statistics\/}~{\em
  44\/}(3), 319--335.

\bibitem[\protect\citeauthoryear{Kristensen, Nielsen, Berg, Skaug, and
  Bell}{Kristensen et~al.}{2016}]{kristensen2016tmb}
Kristensen, K., A.~Nielsen, C.~W. Berg, H.~Skaug, and B.~M. Bell (2016).
\newblock {TMB}: {A}utomatic differentiation and {L}aplace approximation.
\newblock {\em Journal of Statistical Software\/}~{\em 70\/}(i05).

\bibitem[\protect\citeauthoryear{McNeil, Frey, and Embrechts}{McNeil
  et~al.}{2005}]{mcneilquantitative}
McNeil, A., R.~Frey, and P.~Embrechts (2005).
\newblock {\em Quantitative Risk Management: Concepts, Techniques, and Tools}.
\newblock Princeton University Press: Princeton, NJ.

\bibitem[\protect\citeauthoryear{Merton}{Merton}{1976}]{merton1976option}
Merton, R.~C. (1976).
\newblock Option pricing when underlying stock returns are discontinuous.
\newblock {\em Journal of Financial Economics\/}~{\em 3\/}(1), 125--144.

\bibitem[\protect\citeauthoryear{Press, Teukolsky, Vetterling, and
  Flannery}{Press et~al.}{1992}]{pressnumerical}
Press, W.~H., S.~A. Teukolsky, W.~T. Vetterling, and B.~P. Flannery (1992).
\newblock Numerical recipes in {C}.
\newblock {\em Cambridge: Cambridge University\/}.

\bibitem[\protect\citeauthoryear{Preston and Wood}{Preston and
  Wood}{2012}]{preston2012approximation}
Preston, S. and A.~T. Wood (2012).
\newblock Approximation of transition densities of stochastic differential
  equations by saddlepoint methods applied to small-time {Ito}--{Taylor}
  sample-path expansions.
\newblock {\em Statistics and Computing\/}~{\em 22\/}(1), 205--217.

\bibitem[\protect\citeauthoryear{{R Core Team}}{{R Core
  Team}}{2018}]{r2018project}
{R Core Team} (2018).
\newblock {\em R: A Language and Environment for Statistical Computing}.
\newblock Vienna, Austria: R Foundation for Statistical Computing.

\bibitem[\protect\citeauthoryear{Strawderman}{Strawderman}{2004}]{strawderman2004computing}
Strawderman, R.~L. (2004).
\newblock Computing tail probabilities by numerical fourier inversion: The
  absolutely continuous case.
\newblock {\em Statistica Sinica\/}, 175--201.

\bibitem[\protect\citeauthoryear{Tweedie}{Tweedie}{1984}]{tweedie1984index}
Tweedie, M. (1984).
\newblock An index which distinguishes between some important exponential
  families.
\newblock In {\em Statistics: Applications and new directions: Proc. Indian
  statistical institute golden Jubilee International conference}, Volume 579,
  pp.\  6o4.

\bibitem[\protect\citeauthoryear{Wood, Booth, and Butler}{Wood
  et~al.}{1993}]{wood1993saddlepoint}
Wood, A.~T., J.~G. Booth, and R.~W. Butler (1993).
\newblock Saddlepoint approximations to the {CDF} of some statistics with
  nonnormal limit distributions.
\newblock {\em Journal of the American Statistical Association\/}~{\em
  88\/}(422), 680--686.

\bibitem[\protect\citeauthoryear{Yu}{Yu}{2004}]{yu2004empirical}
Yu, J. (2004).
\newblock Empirical characteristic function estimation and its applications.
\newblock {\em Econometric reviews\/}~{\em 23\/}(2), 93--123.

\bibitem[\protect\citeauthoryear{Zhang and Schmidt}{Zhang and
  Schmidt}{2016}]{zhang2016approximation}
Zhang, L. and W.~M. Schmidt (2016).
\newblock An approximation of small-time probability density functions in a
  general jump diffusion model.
\newblock {\em Applied Mathematics and Computation\/}~{\em 273}, 741--758.

\end{thebibliography}
\end{document}